\documentclass[prl,aps,twocolumn]{revtex4}
\usepackage[applemac]{inputenc}
\usepackage{amsmath}
\usepackage{amsfonts}
\usepackage{amssymb}
\usepackage{graphics}
\usepackage{graphicx}
\usepackage{subfigure}
\usepackage{upgreek}
\usepackage{color}

\newcommand{\ii}{\text{i}}

\newcommand{\um}{\mu\text{m}}

\begin{document}

\title{ Enhanced and reduced transmission of acoustic waves with bubble meta-screens}

\author{Alice Bretagne}
\affiliation{ESPCI ParisTech, CNRS, Institut Langevin,
10 rue Vauquelin, 75005 Paris,
France}

\author{Arnaud Tourin}
\affiliation{ESPCI ParisTech, CNRS, Institut Langevin, 
10 rue Vauquelin, 75005 Paris,
France}

\author{Valentin Leroy}
\affiliation{Laboratoire Mati\`ere et Syst\`emes Complexes, Universit\'e Paris Diderot-Paris $7$, CNRS (UMR $7057$), France}

\date{\today}

\begin{abstract}
We present a class of sonic meta-screens for manipulating air-borne acoustic waves at ultrasonic or audible frequencies. Our screens consist of periodic arrangements of air bubbles in water or possibly embedded in a soft elastic matrix. They can be used for soundproofing, but also for exalting transmission at an air/water interface or even to achieve enhanced absorption. 
\end{abstract}
\maketitle

Soundproofing is a challenging task at low frequencies because it involves blocking large-wavelength waves, which requires thick and/or heavy materials. 
To overcome this difficulty composite materials with locally resonant sub-wavelength structural units have been recently developed ~\cite{Yang2010}. Among all possible acoustic sub-wavelength resonators bubbles seem to be very interesting candidates.
Gas bubbles in liquids or soft solids are indeed well known for exhibiting a low frequency resonance, the so-called Minnaert resonance, whose angular frequency is given by $\omega_M = \sqrt{(3\beta_\text{g}+4\mu)/\rho}/a$, where $\beta_\text{g}$ is the longitudinal modulus of the gas, $\mu$ and $\rho$ the shear modulus and mass density of the elastic matrix, and $a$ the radius of the bubble. For a $0.1\,$mm-radius air bubble in water, the resonance frequency is around $30\,$kHz, which corresponds to a wavelength $500$ times larger than the radius of the bubble.
\begin{figure}[h!bt]
\begin{center}
\includegraphics[width = \linewidth]{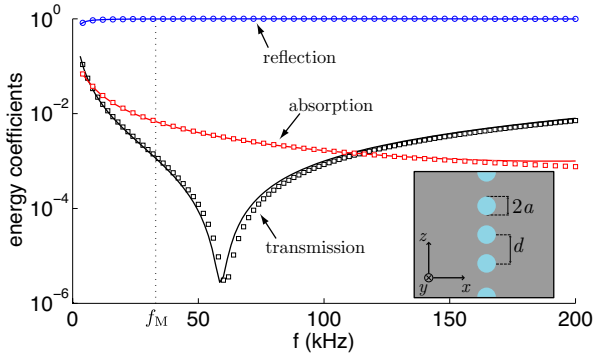}
\vspace{-.5cm}
\caption{Energy transmission, reflection and absorption for a layer of bubbles in water with $a=100\,\um$ and $d/a=5$, as predicted by the SCA (solid lines) and calculated by MST (markers). The inset shows the geometry of the layer.\label{figure1}}
\end{center}
\end{figure}
Recently, experiments have shown~\cite{Leroy2009, Bretagne2009} that a single layer of bubbles was very efficient for blocking low frequency ultrasonic waves in water. An illustration is presented in Fig.~\ref{figure1}, in which the reflection, transmission and absorption coefficients predicted by the multiple scattering theory (MST)~\cite{Sainidou2005}\footnote{Dissipation was included in the MST following the procedure described in~\cite{Leroy2009}.} are shown for a plane wave incoming in the $x$ direction, with wave-number $k$, on a one-layer crystal of bubbles with a squared lattice (see inset). It appears that this bubble meta-screen acts as a nearly perfect mirror, with a minimum of transmission for a frequency close to the resonant frequency of the bubbles ($f_M=\omega_M/2\pi$).



One can use a self-consistent approximation (SCA) to obtain analytical expressions for the amplitude reflection $r$ and transmission $t$, valid  for $kd\ll 1$:~\cite{Leroy2009}
 \begin{subequations}
\begin{eqnarray}
r &=&  \frac{\ii Ka}{\Omega - \ii (\delta + Ka)} \label{eqn2a}\\
t &=& \frac{\Omega - \ii \delta}{\Omega - \ii (\delta + Ka)}, \label{eqn2b}
\end{eqnarray}\label{eq1}
\end{subequations}
where $\Omega = (\omega_M/\omega)^2-(1-2\sqrt{\pi}a/d)$, $K=2\pi/(kd^2)$, and $\delta$ is the dissipative damping of the bubbles.  Eq.~(\ref{eqn2b}) predicts a minimum of transmission at $f_\text{min}=f_M/\sqrt{1-A}$, where $A=2\sqrt{\pi}(a/d)$ is a geometrical factor. 
Note that the viscous and thermal losses $\delta$ are generally much smaller than the super-radiative term $Ka$. 
 For instance, for the meta-screen considered in Fig.~\ref{figure1}, $\delta \simeq 2\times 10^{-2}$ at $50\,$kHz, whereas $Ka\simeq10$. This super-radiative term explains the mirror-like behavior of the bubble meta-screen: Eq.~(\ref{eqn2a}) reduces to $r\simeq -1$ for $\Omega, \delta \ll Ka$. 
  Experiments with bubble meta-screens in soft elastic materials~\cite{Leroy2009, Bretagne2009} and MST calculations~ (see Fig.~\ref{figure1}) have confirmed that the SCA was successful, provided that the bubbles were not too close to each other~\cite{Leroy2009} ($d/a\ge 5$).

\begin{figure}[ht]
\begin{center}
\includegraphics[width =\linewidth]{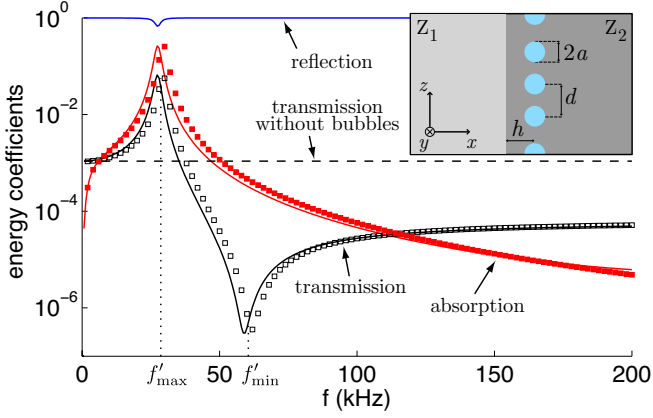}
 \vspace{-.5cm}
\caption{Energy transmission, reflection and absorption for an acoustic wave coming at normal incidence on an air-water interface with one layer of bubbles at distance $h$ from the interface (see inset). MST (markers) and SCA (lines) calculations are shown for $a=100\,\um$,  $d/a=5$, and $h/a=2$. \label{figure2}}
\end{center}
\end{figure}

In this letter, we investigate the ability of a bubble meta-screen to block \emph{air-borne} acoustic waves. The question is whether the minimum of transmission survives when the meta-screen is close to an interface with air. The answer is not straightforward because the presence of an interface is expected to affect the acoustic response of the bubbles, potentially destroying the desired effect. Let us consider the case depicted in the inset of Fig.~\ref{figure2}: the plane wave first goes from a medium with impedance $Z_1$ to a medium with impedance $Z_2$ before it impinges the layer of bubbles. It follows that multiple reflections between the interface and the layer of bubbles occur. Noting $x_{12}=2Z_2/(Z_1+Z_2)$ the transmission from medium $1$ to medium $2$, the total amplitude transmission predicted by the SCA is given by $
t^\prime = x_{12} t/(1-r(x_{21}-1)\text{e}^{2\ii kh})$, where $h$ is the interface-layer distance. In the case of $Z_1\ll Z_2$ (air-water interface, for instance), and if $kh\ll 1$, the energy transmission $T^\prime=(Z_1/Z_2)|t^\prime |^2$ then reduces to:
\begin{equation}
T^\prime =T^\prime_0 \left|\frac{\Omega - \ii \delta}{\Omega-B-\ii (\delta+B kh)}\right|^2, 
\label{eqTprime}
\end{equation}
where $T^\prime_0 = 4Z_1/Z_2$ is the transmission in the absence of bubbles, and $B=4\pi ah/d^2$. Energy reflection and absorption can also be calculated with the same scheme.  Fig.~\ref{figure2} shows that  Eq.~(\ref{eqTprime}) is in a good agreement with the predictions of the MST for the layer of bubbles we considered previously placed at a distance $h=2a$ from an air-water interface. It appears that the presence of the interface does not change the position of the minimum of transmission, for which the sound transmission loss (STL) is increased by $34\,$dB. Note that the reduced transmission with the bubble meta-screen extends over a large range of frequency: even at higher frequencies, the transmission reaches an asymptotic limit which corresponds to a $12\,$dB increase of the STL compared to the bubble free interface.

Most interestingly,  Eq.~(\ref{eqTprime}) also predicts a \textit{maximum} of transmission at $f^\prime_\text{max}=f_M/\sqrt{1-A+B}$.
 It means that the presence of a layer of bubbles close to an interface can \textit{enhance} the transmission of sound. It is tempting to call this effect an extraordinary acoustical transmission, by analogy with the extraordinary optical transmission (EOT) through perforated opaque metallic films~\cite{Ebbesen1998}. However, even though they share some characteristics, both phenomena have different physical origins. In particular, there is no equivalent of the surface-plasmon-polariton modes in the present case. Indeed, the enhanced transmission is simply due to a Fabry-Pérot resonance. The first multiple reflected wave has a phase shift of $\varphi_r + \pi +2kh$ compared with the direct transmitted wave, where $\varphi_r$ is the phase shift induced by a reflection on the bubble meta-screen. Since $\varphi_r \simeq \pi - \Omega/Ka$ (see Eq.~(\ref{eqn2a})), there are constructive interferences at $f^\prime_\text{max}$, leading to the enhanced transmission. 

\begin{figure}[h!bt]
\begin{center}
\includegraphics[width = \linewidth]{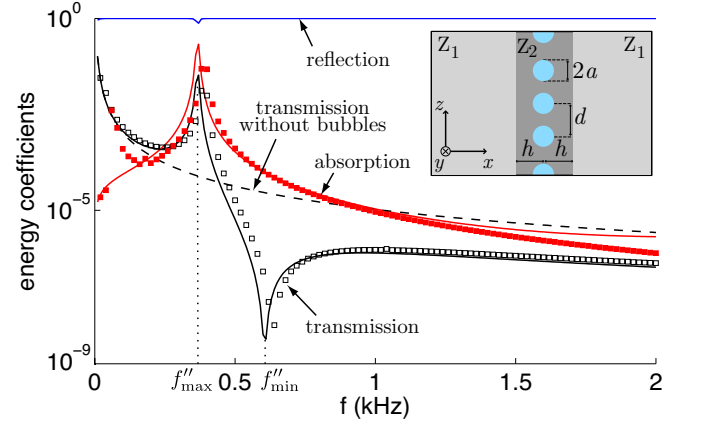}
 \vspace{-.5cm}
\caption{Energy transmission, reflection and absorption for an acoustic wave coming at normal incidence on an air-water-air system with one layer of bubbles (inset). MST (markers) and SCA (lines) calculations are shown for $a=1\,$cm,  $d/a=5$, and $h/a=2$.\label{figure4}}
\end{center}
\end{figure}

When a second interface is introduced (see the inset of  Fig.~\ref{figure4}), even more multiple reflections are involved in the transmission. The algebra is still tractable and, in the $Z_1\ll Z_2$ and $kh\ll 1$ limit, the SCA predicts an energy transmission given by
\begin{equation}
T^{\prime\prime} =T^{\prime\prime}_0 \left|\frac{\Omega - \ii \delta}{\Omega-B/2-\ii( \delta+2\Omega kh)}\right|^2, \label{eqTsec}
\end{equation}
where $T^{\prime\prime}_0 = Z_1^2/(\rho_2 \omega h)^2$ is the usual mass density law for transmission through a high impedance thin wall ($\rho_2$ is the mass density of medium $2$).  As in the one interface case, Eq.~(\ref{eqTsec}) predicts the existence of a minimum and maximum of transmission. To emphasize the possible applications for soundproofing, we show an example in the audible frequency range. As the Minnaert resonance scales as the inverse of the bubble radius, $100$ times larger bubbles are required to shift the $60\,$kHz seen in Fig.~\ref{figure2} to $600\,$Hz, which makes a radius of $1\,$cm. Keeping the same $d/a$ and $h/a$ ratios as in Fig~\ref{figure2}, we thus consider a $4\,$cm-thick wall of water
embedded with a layer of $1\,$cm radius bubbles separated by a $5\,$cm lattice constant (see Fig.~\ref{figure4}).  Agreement with the calculations by MST is satisfactory. The minimum of transmission is at $600\,$Hz, as predicted, and it corresponds to an extra $35\,$dB of transmission loss compared to the mass density law. 
 Besides, the STL is not increased only at resonance: for higher frequencies, bubbles bring on average $10\,$dB to the loss. 

Bubble meta-screens thus appear to be efficient systems for blocking air-borne acoustic waves: the minimum of transmission survives to the presence of one or two air interfaces. More surprisingly, a bubble meta-screen close to an interface can also enhance the transmission, which could be useful for impedance matching applications. The question of the feasibility naturally arises. Water walls with arrays of bubbles are of course unrealistic. Nevertheless, it has been confirmed by experiments~\cite{Bretagne2009} that bubbles can be embedded in a soft elastic matrix without losing their low frequency resonance, which means that the practical realization of elastic meta-screens is possible. Analytical expressions for the transmission, reflection and absorption coefficients are then a powerful tool for designing meta-screens. We report in Table~\ref{tab1} the equations that govern the main features of the transmission. They can be used as guidelines for the choice of parameters $a$, $d$ and $h$ for a bubble meta-screen. Let us consider for instance a soundproofing application. First, $a$ should be chosen so that the resonance frequency of the bubbles is close to the frequency one needs to block. Then, the choice of $d$ is a trade-off: a small $d$ makes the minimum of transmission deeper but it shifts it to a higher frequency, which is not desirable for low frequency soundproofing. On the other hand, $h$ should be taken as large as allowed by the practical application (and within the $kh\ll 1$ limit): a large $h$ deepens the minimum of transmission, and it shifts the maximum of transmission to a lower frequency. We recall that the SCA is valid only for $d/a\ge 5$. Furthermore, because of limitations of the MST program we used, we were not able to test its validity for $h/d< 0.4$.

\begin{table}[htdp]
\caption{Equations for the positions and values of the minimal, maximal and asymptotic transmissions as predicted by the SCA. Factors $A=2\sqrt{\pi}a/d$ and $B=4\pi ah/d^2$ depend only on the geometry of the device.}
 \vspace{-.5cm}
\begin{center}
\begin{tabular}{|l c c c|}
\hline
  & no interface & $1$ interface & $2$ interfaces  \\
  \hline\hline
  $f_\text{max}$		& /				& $f_M/\sqrt{1-A+B}$& $f_M/\sqrt{1-A+B/2}$\\
\hline
 $T_\text{max}/T_0$& /				& $[\delta/B+kh]^{-2}$	& $(1/4)[\delta/B+kh]^{-2}$ \\
\hline
 $f_\text{min}$ 		& \multicolumn{3}{c|}{$f_M/\sqrt{1-A}$} \\
\hline
 $T_\text{min}/T_0$ 	&  $(\delta/Ka)^2$	& $(\delta/B)^2$	& $4(\delta/B)^2$ \\
\hline
$T_\text{asy}/T_0$ 	&  /	& $[1+B/(1-A)]^{-2}$	& $[1+B/2(1-A)]^{-2}$ \\
\hline
\end{tabular}
\end{center}
\label{tab1}
\end{table}

\begin{figure}[tbh]
\begin{center}
\includegraphics[width = .9\linewidth]{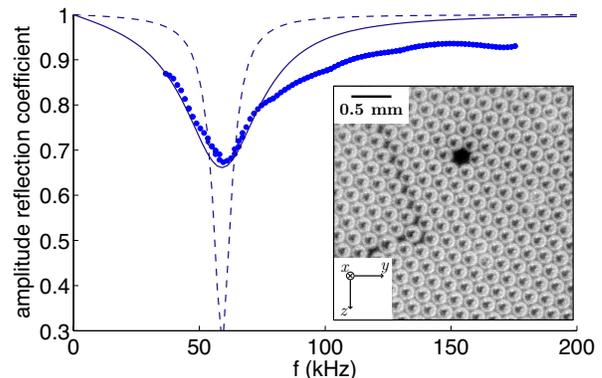}
 \vspace{-.5cm}
\caption{Inset: top view of a bubble raft at the surface of a surfactant solution. Monodisperse small bubbles are injected by blowing air through a $20\,\um$-diameter capillary. Reflection measurements are performed by acoustic pulses emission and reception with a couple of broadband air transducers. Note that, due to the size of the transducers, the incindence was not normal but set to a minimal value of $20^\circ$. Plots: markers are the experimental amplitude reflection coefficient for $90\,\um$-radius bubbles, compared with the MST calculation for bubbles in water (dashed line) and in a $250$ times more viscous fluid (solid line).\label{figure3}}
\end{center}
\end{figure}

However, MST calculations show that more compact bubble meta-screens qualitatively behave the same way: they still exhibit both a maximum and a minimum of transmission. We checked this point experimentally with the most compact structure one can imagine: a bubble raft~\cite{Bragg1947} (see inset of  Fig.~\ref{figure3}). 
As shown in Fig.~\ref{figure2}, the enhanced transmission is expected to be accompanied with a maximum of absorption and a minimum of reflection. 
Fig.~\ref{figure3} reports the reflection coefficient we measured for a raft of $90\,\um$-radius bubbles. A minimum is clearly seen at $60\,$kHz, a position perfectly predicted by the MST calculation. Experiments with different bubble sizes confirmed that this position scaled as the inverse bubble radius, as expected. The depth of the minimum is overestimated (dashed line in Fig.~\ref{figure3}) by the MST. This discrepancy may be attributed to a higher effective viscosity of the liquid, due to its confinement between bubbles; an hypothesis which is supported by a better agreement when the viscosity is increased by a factor $250$ (solid line in Fig.~\ref{figure3}). A side effect of this high effective viscosity is a poor enhancement of the transmission: MST calculation shows that, at normal incidence, only $0.3\,\%$ of the energy is transmitted (below our experimental sensitivity). However, absorption is significantly increased: $54.8\,\%$ of the energy is absorbed. It indicates that compact bubble meta-screens, such as bubble rafts, might be used as absorbing materials.

\end{document}